# Magnetic hyperthermia in single-domain monodisperse FeCo nanoparticles: Evidences for Stoner-Wohlfarth behaviour and large losses

L.-M. Lacroix[1], R. Bel Malaki[1], J. Carrey[1,*], S. Lachaize[1], G.F. Goya[2], B. Chaudret[3], M. Respaud[1]

[1]Université de Toulouse; INSA; UPS; LPCNO, 135 avenue de Rangueil, F-31077 Toulouse, France and
CNRS; LPCNO, F-31077 Toulouse, France

[2]Instituto de Nanociencia de Aragón (INA), Universidad de Zaragoza, Interfacultades II - Pedro Cerbuna 12, 50009-Zaragoza, Spain

[3]Laboratoire de Chimie de Coordination-CNRS, 205 rte de Narbonne, 31077 Toulouse cedex 4, France

**Abstract :**

We report on hyperthermia measurements on a colloidal solution of 14.2±1.5 nm monodisperse FeCo nanoparticles (NPs). Losses as a function of the magnetic field display a sharp increase followed by a plateau, which is what is expected for losses of ferromagnetic single-domain NPs. The frequency dependence of the coercive field is deduced from hyperthermia measurement and is in quantitative agreement with a simple model of non-interacting NPs. The measured losses (1.5 mJ/g) compare to the highest of the literature, though the saturation magnetization of the NPs is well below the bulk one.

**Main Text:**

I. Introduction

Hyperthermia in oncology consists in rising the temperature of a cancerous tumour to improve the efficiency of chemotherapy and radiotherapy. For this purpose, one promising technique is based on the use of magnetic nanoparticles [1,2,3]. In a therapeutic process, magnetic nanoparticles (NPs) must be first accumulated inside the tumour and then excited by an alternating magnetic field. Optimal frequency $f$ and magnetic field $H_{max}$ values result from a compromise between the heating efficiency and the necessity to avoid harmful effects to the patient, which are minimized if the product $H_{max}.f$ does not exceed $5\times10^9$ Am$^{-1}$s$^{-1}$ [3]. Active research is done to improve the specific absorption rate (SAR) of magnetic NPs, which could permit the treatment of smaller size tumours [3]. Regarding their natural biocompatibility and their numerous synthesis procedures, the iron oxides are the most studied materials. The highest SAR values for these materials are 600 W/g (400 kHz, 11 kA/m) for chemically-synthesized NPs [4] and 960 W/g (410 kHz, 10 kA/m) for NPs synthesized by bacteria [5]. With respect to metallic NPs, SAR values of 500 W/g (400 kHz, 13 kA/m) and 720 W/g (410 kHz, 10 kA/m) were reported for Co NPs [1,6].

The SAR amplitude is given by SAR= $A.f$, where $A$ is the specific area of the hysteresis loop (i.e. specific losses) at the frequency and magnetic field at which the experiment is conducted. Given the experimental SAR values given above, the largest losses values reported in literature for chemically-prepared NPs are in the range 1.25-1.75 mJ/g. SAR amplitude versus the external parameters strongly depends on the NP dynamic magnetic properties. First, when the particles are large enough to be multi-domain, the reversal of the magnetization occurs by wall motion, leading to moderate heating and to a cubic dependence of SAR versus $H_{max}$ at low field [1]. Single-domain NPs have more interesting properties for magnetic hyperthermia [7]. The process for their magnetization reversal is similar to a macrospin, as described by Stoner and Wohlfarth [8]. It consists in the thermally activated jumps over an energy barrier, the height of which depends on the anisotropy $K$, on the volume $V$ of the NP, and on the applied magnetic field. The hysteresis loop displays a complex dependence on $f$, $H_{max}$, $K$, $V$, and on the temperature $T$, which cannot be derived analytically [9,10]. The hysteresis loop evolves progressively from a Langevin function for a zero frequency – or equivalently an infinite temperature - to a Stoner-Wohlfarth (SW) hysteresis at infinite frequency or null temperature. To calculate $A$ and interpret hyperthermia experiments, two models -valid in different regimes- can be used. First, when the applied magnetic field is small compared to the saturation field of the NPs, the linear response theory can be used. In this case the hysteresis loop is an ellipse of area

$$A = \frac{\pi \mu_0 H_{max}^2}{\rho} \chi''(f) \qquad (1)$$

with

$$\chi'' = \frac{\mu_0 M_S^2 V}{3 k_B T} \left( \frac{f \tau_N}{1 + (f \tau_N)^2} \right). \qquad (2)$$

$M_s$ is the saturation magnetization, $\rho$ the density of the material, $\tau_N = \tau_0 \exp\left(\frac{KV}{k_B T}\right)$ the Néel relaxation time, and $\tau_0$ the interwell relaxation time. The linear theory is suitable to interpret hyperthermia experiments on superparamagnetic NPs, since the rather weak magnetic field used (generally up to 30 mT) is far from the saturation field of the NPs. In this case, SAR is proportional to $H_{max}^2$ and its amplitude can be deduced directly from a measurement of the imaginary part of the susceptibility $\chi''$ (see Eq. (1)). These dependences have been often verified in the literature [4,6,11]. However, the linear theory is inappropriate for ferromagnetic NPs or for NPs close to the superparamagnetic/ferromagnetic transition, since the applied magnetic field can be close to the saturation field. In the ferromagnetic state, the SW model is more suitable to interpret the experimental data. The classical SW model is strictly valid for an infinite frequency or at $T = 0$ but indicates the general features of the expected properties: at $T = 0$, the area as a function of the magnetic field presents a sharp increase at the coercive field followed by a plateau

[3]. This plateau corresponds to the largest hysteresis area possible for single-domain NPs. For randomly oriented uniaxial NPs, this area is $A = \alpha.\mu_0.M_s.H_{C0}$, where $\alpha \approx 2$ and $H_{C0}$ is the value of the coercive field $H_C$ at $T = 0$. Because $H_C$ and $A$ both vary with temperature and frequency [10,12], the prediction of the SAR remains an open problem, since they cannot be estimated through simple extrapolation of the static and low-frequency conventional measurements [1,13].

The previous considerations explain why NPs in the SW regime are highly desirable for hyperthermia applications; to get the largest SAR possible, the NPs should be saturated by $H_{max}$, with a coercive field as large as possible but below $H_{max}$. Given the weak magnetic field used in hyperthermia, this is only possible for soft ferromagnetic NPs or super-paramagnetic NPs close to the ferromagnetic transition. To synthesize such NPs and characterize them for hyperthermia applications, several requirements should be fulfilled : i) particles should be single-domain and rather monodisperse to avoid convolution effects or the presence of too many multi-domain NPs ii) they should have a small enough coercive field even at high frequency so that a moderate magnetic field can saturate their magnetization iii) measurements should be performed on systems where dipolar interactions do not influence too much the NP intrinsic magnetic properties. The meeting of these three requirements at the same time is not trivial. Especially, in the size range of interest 10-40 nm, many synthesis methods fail in reaching a narrow size distribution. This may explain why, so far, experimental evidences for SW behaviour in hyperthermia measurements have never been reported.

In this Letter, we present hyperthermia measurements on a colloidal solution of 14.2±1.5 nm monodisperse FeCo NPs. For the first time, clear features expected for NPs in the SW regime are observed in hyperthermia measurements. A variation of the coercive field of the NPs as a function of frequency is deduced and quantitatively analysed in the framework of the SW model. Moreover the losses compare to the highest reported in the literature for chemically-synthesized NPs, whereas the NP magnetization is well below the bulk one.

To obtain monodisperse FeCo NPs, we use an organometallic synthesis leading to super-lattices of NPs, which was described elsewhere [14]. The colloidal solution is then obtained by a sonication-assisted dispersion of the super-lattices in THF solvent with octylamin. The colloidal solution on which the measurements and characterisations reported in this article have been performed contains 63 mg of $Fe_{0.63}Co_{0.37}$ alloy, 104 mg of surfactants and 822 mg of THF, leading to a metal concentration of 0.8 % vol. A transmission electron microscopy micrograph of the NPs is shown in Fig. 1(a). The mean diameter of the NPs is 14.2 nm, with a standard deviation of 1.5 nm. Particle size analysis of the colloidal solution by dynamic light scattering displays a small peak at a diameter of 19 nm, corresponding to isolated NPs, and a large amplitude broad peak between 100 nm and 1 µm, corresponding to agglomerates of several NPs [see Fig. 1(b)]. $M_S$ measured by SQUID on powder is 144 $Am^2.kg^{-1}$, well below the bulk value (240 $Am^2kg^{-1}$), due to an imperfect alloying of Fe and Co atoms [14]. On the powder, the remnant magnetization $M_R$ at 2 K is almost null, due to strong dipolar interactions between NPs. In Fig. 2, SQUID measurement of the colloidal solution at 2 K is displayed. The ratio $M_R/M_S$ at 2 K is about 0.4, just below the expected value of 0.5 for magnetically independent NPs. This indicates that the NPs are still partly magnetically coupled in the colloidal solution, due to the presence of agglomerates. The corresponding coercive field is $\mu_0H_C = 32$ mT. Hyperthermia experiments were performed on a home-made frequency-adjustable electromagnet, which was described elsewhere [15]. For this purpose, a vessel containing the colloidal solution was sealed under vaccum to prevent any oxidation of the NPs. The vessel is then placed in a calorimeter with 2 ml of water, the temperature of which is measured.

Fig. 3(a) displays typical experiments showing the evolution of the temperature as a function of time [$T(t)$] after switching on the magnetic field. In the first two minutes, non-reproducible oscillations due to convection-conduction phenomena inside the calorimeter are observed, causing large uncertainties in the determination of the initial slope of the $T(t)$, especially for large SARs. A reproducible SAR value is determined from the mean slope of the $T(t)$ functions between 0 and 200 s. SAR values calculated this way are underestimated by approximately 20 % since we neglect the losses of the calorimeter, which are around $3\times10^{-2}$ W.K$^{-1}$. The uncorrected values are used in this article. Dividing SAR values by the magnetic field frequency leads to the specific losses $A$. Losses as a function of magnetic field, measured at frequencies in the range 2-100 kHz, are summarized in Fig. 3(b). They display a sharp increase at a given magnetic field and progressively saturate at high field, which is precisely the behaviour expected for SW NPs. We identify the coercive field $H_C$ as the field at which the slope of the SAR($\mu_0 H$) is maximum. It evolves from $\mu_0 H_C = 7$ mT at $f = 2$ kHz to $\mu_0 H_C = 12$ mT at $f = 100$ kHz [see Fig. 3(d)].

The frequency dependence of the SAR [SAR($f$)] was more specifically studied for four different values of the magnetic field ranging from $\mu_0 H = 3.87$ mT to $\mu_0 H = 29$ mT [see Fig. 3(c)]. Except for $\mu_0 H = 10.65$ mT, SAR is simply proportional to the frequency. In the range between 6 and 12 mT, it is expected from Fig. 3b that the frequency dependence of SAR may not be linear because of the frequency dependence of $H_C$. This is what is observed in experiments performed at $\mu_0 H = 10.65$ mT, for which the slope of the SAR($f$) is much higher at low frequency. For $\mu_0 H = 29$ mT, a linear fit of the SAR($f$) allows us to determine the losses per cycle at saturation to 1.5 mJ/g, in agreement with Fig. 3b. This value is similar to the ones reported for the best chemically-prepared iron oxides NPs (1.5 mJ/g [16]) and Co NPs (1.25 mJ/g [1] and 1.75 mJ/g [17]).

We will now quantitatively analyse the increase of coercive field with frequency visible in Fig. 3(b), which is summarized in Fig. 3(d). As seen in the particle-size analysis and the magnetic measurements, our system contains magnetically coupled nanoparticles, so these data should in principle be analysed in the framework of a SW model with interactions. However, the frequency dependence of the coercive field has never been studied in this case due to the complexity of matching the "time" of the Monte-Carlo simulations used to study these systems to a real time [18]. As a consequence, there is so far no theoretical or phenomenological expression adapted to this problem. However, we will show that a simple model of non-interacting SW NPs leads to coherent results in our case.

In an assembly of independent randomly oriented single-domain uniaxial NPs, the temperature and frequency variation of the coercive field is described by the following equation [12,19]:

$$\mu_0 H_C = \frac{2K}{\rho M_S}\left[0.479 - 0.81\left(\frac{k_B T}{2KV}\left(\ln 1/f\tau_0\right)\right)^{\frac{3}{4}}\right], \quad (3)$$

where $\rho = 8300$ kg.m$^{-3}$ is the density of the material. $f$ represents the inverse of the typical measurement time, which will be stated here as the time of a complete hysteresis cycle. This equation is only valid if the NPs can be considered as fixed in the liquid medium, i.e. if the measurement frequency $f$ is much greater than the Brownian frequency $f_B$ of the NPs. Dynamic light scattering measurements on the colloidal solution shows that 95 % of the NPs are contained in agglomerates of hydrodynamic radii $r$ larger than 350 nm. The Brownian frequency $f_B$ of these agglomerates, calculated using $f_B = k_B T / 4\pi\eta r^3$, where $\eta=0.48\times10^{-3}$ kg.s$^{-1}$m$^{-1}$ is the viscosity coefficient of the solvent leads to $f_B$ = 130 Hz. Thus Eq. (3) can safely be used to analyse our experimental data obtained in the range 2-100 kHz. Using $M_S$ and $V$ values deduced from magnetic and microscopy measurements and $\tau_0 = 10^{-9}$ s, the best fit of experimental data is obtained for $K = 4.5 \times 10^4$ J/m$^3$ [see Fig. 3(d)]. Another value of the anisotropy can be determined using the coercive field $\mu_0 H_C$ =32 mT obtained by the SQUID measurement at 2 K (see Fig. 2). Using Eq. (3) and taking $f = 5.4\times10^{-5}$ Hz leads to $K = 4.2 \times 10^4$ J/m$^3$, in good agreement with the value determined from hyperthermia experiments.

Finally, we present another way to estimate the losses of the NPs. The hysteresis area for independent randomly oriented SW NPs is given by $A \approx 2\mu_0.M_s.H_C$ [12]. This calculation leads to losses varying from 1.4 mJ/g at 2 kHz to 3.4 mJ/g at 100 kHz. At low frequency, the value determined by this method is in agreement with the losses determined calorimetrically [1.5 mJ/g, see Fig. 3(c)]. However, we did not observe the expected increase of losses with frequency at high magnetic field but rather a frequency independent value [see Fig. 3(c)]. The origin for this is so far not fully understood, but two hypotheses can be put forward to explain it. First, one could think that the NPs could be unsaturated at high frequency due to an increase of the saturation field. The shapes of the experimental curves in Fig. 3(b) do not seem to support this hypothesis, but experiments at higher magnetic field than the one accessible on our bench could be interesting to validate our validate it. Second, it is also conceivable that the magnetic interactions between the NPs could lead to a non-trivial frequency-dependence of the hysteresis area. Theoretical studies about the frequency variation of hysteresis shape and area in the case of interacting SW NPs are still lacking to go further into this discussion.

As a conclusion, we report a clear signature of SW behaviour in magnetic hyperthermia experiments. The magnetic field dependence of the losses follows the theoretically expected step-like shape. A preliminary analysis of the data with the available expressions neglecting the dipolar interactions effects allows us to reproduce quantitatively the frequency-dependence of the coercive field. However, these simple models fail to predict the frequency-dependence of the losses at high magnetic field. We show in this work that using single-domain NPs with a moderate saturation field maximizes the losses. Indeed, values comparable to the highest of the literature are obtained, despite the magnetization of the NPs is well below the bulk one. Losses could be further enhanced by using particles with bulk magnetization. The measurement of large $M_S$ NPs will constitute future developments of this work.

**Acknowledgements :**
We acknowledge C. Crouzet for her precious help on the capacitor box, A. Mari for magnetic measurements, and InNaBioSanté foundation for financial support.

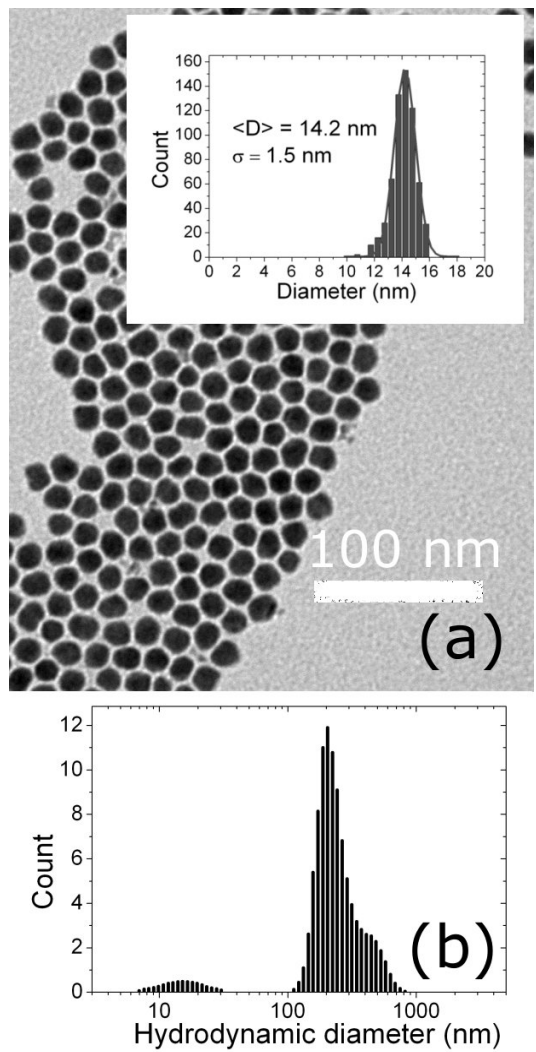

Figure 1 : (a) Transmission electron microscopy of the FeCo NPs measured in hyperthermia. (inset) The size distribution extracted from several images leads to a mean diameter of 14.2 nm and a square deviation of 1.5 nm. (b) Dynamic light scattering experiments of the colloidal solution displaying the hydrodynamic diameter of the aggregates.

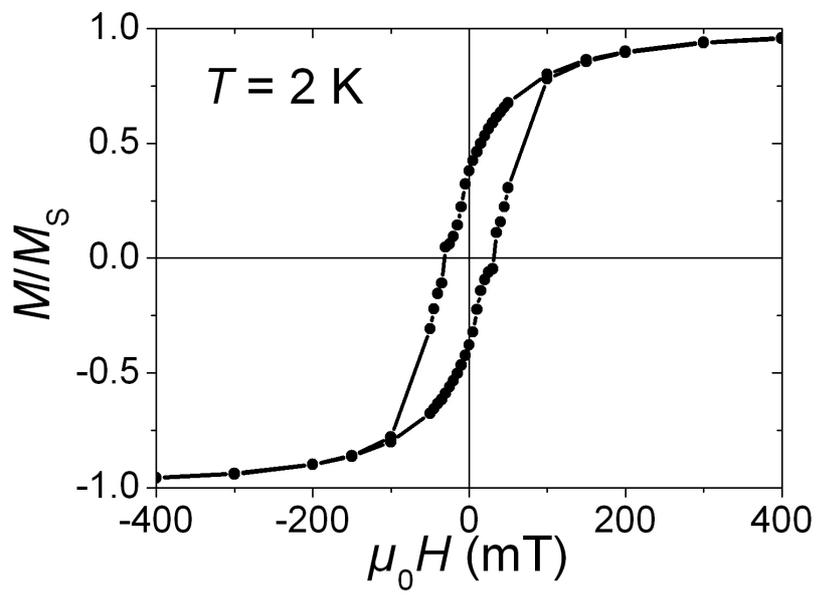

Figure 2 : Magnetic hysteresis loop of the colloidal solution at $T = 2$ K.

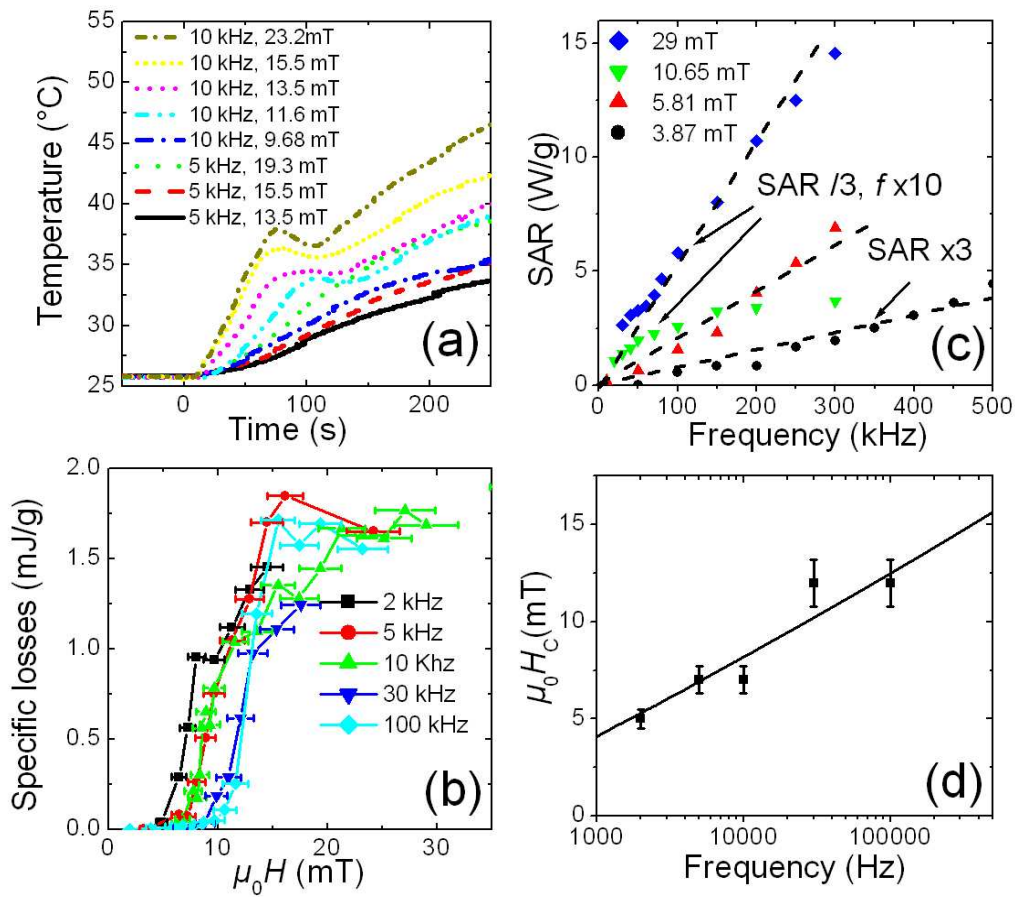

Figure 3 : (a) Evolution of temperature as a function of time after application of the a.c. magnetic field at $t = 0$. (b) Losses per cycle as a function of magnetic field, measured at various frequencies. They are calculated from the mean slope of the $T(t)$ functions between 0 and 200 s. (c) Evolution of the SAR as a function of frequency for various values of the magnetic field. For experiments performed at $\mu_0 H = 10.65$ mT and $\mu_0 H = 29$ mT, SAR values have been divided by 3 and 10, respectively, and frequency has been multiplied by 10. For experiments performed at $\mu_0 H = 3.87$ mT, SAR values have been multiplied by 3. (d) Evolution of the coercive field with frequency, fitted using Eq. (3) with $K = 4.5 \times 10^4$ J/m$^3$.